\pdfoutput=1

\documentclass[aip,reprint,jcp]{revtex4-1}

\usepackage{graphicx}
\usepackage{mathrsfs}

\usepackage{physics} 

\usepackage{siunitx}

\usepackage{xr}
\usepackage{pdfpages}

\usepackage{hyperref}

\makeatletter
\AtBeginDocument{\let\LS@rot\@undefined}
\makeatother

\usepackage{footmisc} 

\begin{document}

{
\makeatletter
\def\frontmatter@thefootnote{%
 \altaffilletter@sw{\@fnsymbol}{\@fnsymbol}{\csname c@\@mpfn\endcsname}%
}%
\makeatother

\title{A Quantum Algorithm for Non-Markovian Electron Transfer Dynamics Using 
Repeated Interactions} 
\title{Repeated Interaction Scheme for the Quantum Simulation of Non-Markovian Electron Transfer Dynamics}

\author{Lea K.\ Northcote}
\thanks{Contributed equally}
\affiliation{NNF Quantum Computing Programme,
Niels Bohr Institute, University of Copenhagen, DK-2100 Copenhagen \O, Denmark}
\author{Matthew S.\ Teynor}
\thanks{Contributed equally}
\affiliation{NNF Quantum Computing Programme,
Niels Bohr Institute, University of Copenhagen, DK-2100 Copenhagen \O, Denmark}
\affiliation{Nano-Science Center and Department of Chemistry, University of Copenhagen, DK-2100 Copenhagen \O, Denmark}
\author{Gemma C.\ Solomon}
\email[Email: ]{gsolomon@chem.ku.dk}
\affiliation{NNF Quantum Computing Programme,
Niels Bohr Institute, University of Copenhagen, DK-2100 Copenhagen \O, Denmark}
\affiliation{Nano-Science Center and Department of Chemistry, University of Copenhagen, DK-2100 Copenhagen \O, Denmark}

\date{1 May 2025}

\begin{abstract}
Quantum algorithms have the potential to revolutionize our understanding of open quantum systems in chemistry. In this work, we demonstrate that a repeated interaction model, which could serve as the foundation for a digital quantum algorithm, can effectively reproduce non-Markovian electron transfer dynamics under four different donor-acceptor parameter regimes and for a donor-bridge-acceptor system. We systematically explore how the model scales for the regimes. Notably, our approach exhibits favorable scaling in the required repeated interaction duration as the electronic coupling, temperature, damping rate, and system size increase. Furthermore, a single Trotter step per repeated interaction leads to an acceptably small error, and high-fidelity initial states can be prepared with a short time evolution. This efficiency highlights the potential of the model for tackling increasingly complex systems. When fault-tolerant quantum hardware becomes available, algorithms based on this model could be extended to incorporate structured baths, additional energy levels, or more intricate coupling schemes, enabling the simulation of real-world open quantum systems that remain beyond the reach of classical computation.
\end{abstract}


\maketitle 



\section{Introduction}
Quantum algorithm research for chemistry has often concentrated on calculating static properties for complex electronic structure problems; however, simulations of chemical dynamics, especially open system dynamics, would also benefit from a level of accuracy that classical computation may not be able to achieve.\cite{cao_quantum_2019,daley_practical_2022} Open quantum systems interact with the environment and the resulting decoherence and relaxation can be challenging to capture fully with classical computation. This challenge becomes even larger for systems that strongly couple to the environment and have non-Markovian effects.\cite{li_non-markovian_2019,li_concepts_2018} For classical simulations of open quantum systems, storing the density matrix usually requires memory that scales exponentially with the size of the system, which limits the systems that can be studied and the methods that can be applied. Because quantum computers encode information in qubits, we can circumvent the exponential scaling issues encountered when encoding quantum states with classical bits.\cite{nielsen_quantum_2010}

Analog quantum simulators are a promising avenue for modeling chemically relevant open quantum systems. For an analog simulation, a quantum device is engineered so that the non-unitary dynamics of a chemical system can be mapped directly onto the dynamics of the device. Studies have proposed strategies for performing analog simulations of molecular open quantum systems on superconducting,\cite{mostame_quantum_2012, garcia-alvarez_quantum_2016} trapped ion,\cite{lemmer_trapped-ion_2018, macdonell_analog_2021, schlawin_continuously_2021} semiconductor quantum dot,\cite{kim_analog_2022} and photonic\cite{sparrow_simulating_2018} quantum devices. With the improved quality of quantum hardware, we have begun to see experimental realizations of these analog simulations. So et al. simulated charge transfer dynamics across multiple parameter regimes on a trapped ion device.\cite{so_trapped-ion_2024} Sun et al. implemented a method for structured baths on trapped ion hardware to simulate vibration-assisted energy transfer.\cite{sun_quantum_2024} Also on a trapped ion device, Navickas et al. recently performed open-system molecular dynamics simulations of photoexcited pyrazine.\cite{navickas_experimental_2024}. In addition to these trapped ion results, there have been experimental realizations using photonics\cite{tang_simulating_2024} and nuclear magnetic resonance\cite{wang_efficient_2018}. While these approaches are promising, analog quantum computation is fundamentally constrained to certain parameter regimes dictated by the physics of the underlying device. In contrast, digital quantum algorithms can reach arbitrary parameter regimes, under the assumption that there will be fault-tolerant hardware with large numbers of qubits that can reach large circuit depths. Therefore, while analog simulations may provide new insights into certain open quantum systems, digital approaches promise broader applicability in the future fault-tolerant quantum computing era.

It is not straightforward to encode the non-unitary dynamics of open quantum systems in a digital quantum algorithm because the building blocks of digital quantum computing, quantum gates, are unitary. A general strategy to overcome this challenge is to introduce ancilla qubits and embed non-unitary operators within higher-dimensional unitary gates. Various methods building on this strategy have been developed, including dilation or block-encoding methods,\cite{hu_quantum_2020, schlimgen_quantum_2021, suri_two-unitary_2023, schlimgen_quantum_2022, gaikwad_simulating_2022, ding_simulating_2024, basile_quantum_2024, xuereb_deterministic_2023} quantum imaginary time evolution (QITE),\cite{kamakari_digital_2022} variational approaches,\cite{endo_variational_2020, shivpuje_designing_2024, watad_variational_2024, luo_variational_2024, mahdian_hybrid_2020, liu_variational_2021, lau_convex_2023, joo_commutation_2023, suri_speeding_2018, santos_low-rank_2025, lee_neural-network_2021, ollitrault_quantum_2023, gravina_adaptive_2024, schlegel_coherent-state_2023, zhou_hybrid_2023} and tensor train approaches.\cite{lyu_tensor-train_2023} Notably, dilation methods based on singular value decomposition (SVD) have shown significant promise for simulating chemically relevant dynamics on near-term quantum hardware. For instance, Hu et al. applied SVD techniques to simulate Lindblad dynamics of the Fenna-Matthews-Olson (FMO) complex, a photosynthetic antenna that channels excitation energy to the reaction center in photosynthesis.\cite{hu_general_2022} Dan et al. also used SVD to implement a quantum algorithm based on the hierarchical equations of motion to model the non-Markovian dynamics of FMO and charge and energy transport in molecular systems.\cite{dan_simulating_2025} These studies are promising, but SVD methods require diagonalization of the propagator, which may limit how the method can be applied in the future.

Recently, quantum algorithms that rely on repeated interaction models have demonstrated considerable potential.\cite{pocrnic_quantum_2024, gallina_strategies_2022, gallina_simulating_2024} These methods use mid-circuit measurements on ancilla qubits to introduce non-unitary effects, see Figure \ref{fig:intro_oqs-ri}. In this framework, the system of interest is coupled to an ancilla qubit register, and the combined system repeatedly undergoes a short unitary evolution before the ancilla is measured and reset to an initial state. The joint unitary evolution and the initial state of the ancilla can be designed to reproduce both Markovian and non-Markovian dynamics of the system of interest.\cite{ciccarello_quantum_2022} Pocrnic et al. gave an analytic proof that quantum circuits based on repeated interaction models can replicate Lindblad dynamics to a given error, where the upper bound for the required number of mid-circuit measurements is inversely proportional to the error, meaning that smaller errors can be achieved by increasing the rate of mid-circuit measurements.\cite{pocrnic_quantum_2024} Gallina et al. used a repeated interaction scheme to simulate energy transfer in a two-chromophore system, focusing on how to calculate response functions with a quantum algorithm.\cite{gallina_simulating_2024} While these advancements are promising, the practical utility of simulating chemical systems with repeated interactions remains underexplored. Specifically, previous studies have either focused on proving worst-case scenario error bounds or applying the method to a specific system, leaving a gap in understanding how these algorithms scale for different chemical regimes.

\begin{figure}[ht]
    \centering
    \includegraphics{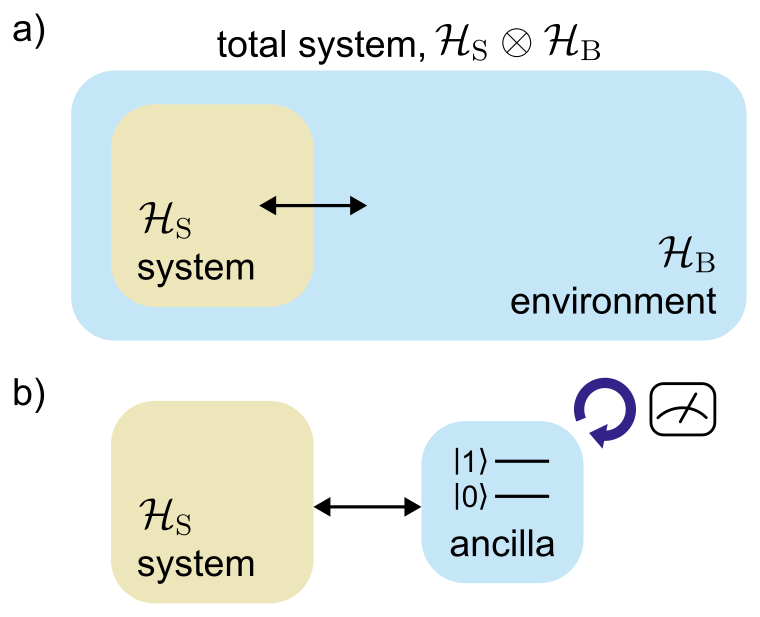}
        \caption{\textbf{Open Quantum Systems and the Repeated Interaction Model.} (a) Illustration of an open quantum system. The system resides in the Hilbert space $\mathcal{H}_\mathrm{S}$ and interacts with the surrounding environment described by $\mathcal{H}_\mathrm{B}$ to give the total system that exists in the combined Hilbert space. (b) Schematic of the repeated interaction model. The effect of the environment on the system is captured by having the system interact with an ancilla qubit that is repeatedly reset with mid-circuit measurements. The ancilla qubit and the measurements allow for the simulation of non-unitary system dynamics.}
        \label{fig:intro_oqs-ri}
\end{figure}

Electron transfer (ET) reactions offer an ideal test case for investigating how the repeated interaction method scales for chemical systems. ET reactions involve the transfer of an electron from a donor to an acceptor and are fundamental across chemistry, from photosynthesis to catalysis. Compared to other model systems, such as energy transfer, ET reactions typically have stronger coupling between the electronic and nuclear degrees of freedom. This strong vibronic coupling arises from a significant reorganization of the nuclear coordinates as the electronic state changes. Due to this strong coupling, ET dynamics can exhibit non-Markovian effects, where the nuclear degrees of freedom do not act as a memoryless bath. To account for this, foundational models for ET reactions explicitly treat both the electronic system and a collective reaction coordinate.\cite{nitzan_chemical_2006} The reaction coordinate can interact with the environment, making reactions irreversible and necessitating an open quantum system treatment of the combined electron and reaction coordinate supersystem. ET reactions are appealing for testing the repeated interaction scheme because the dynamics are complicated, but well-understood, for basic models. So even though we are unlikely to learn new chemistry by simulating simple ET models, these models give us a measure to evaluate when repeated interactions will give chemically meaningful results. Additionally, relevant ET reactions span many parameter regimes,\cite{may_kuhn_2011} which allows for systematic testing of repeated interaction scaling across a large parameter range. For these reasons, we chose to investigate how repeated interaction models scale for ET reactions.

In this paper, we aim to connect theoretical error bounds and practical applicability by examining the scaling behavior of the repeated interaction scheme for ET reaction dynamics. Specifically, we focus on donor-acceptor (DA) systems to investigate how this scheme performs as the electronic coupling, the temperature, and the damping rate increase. Furthermore, we extend the model to a donor-bridge-acceptor (DBA) system to investigate the effect of increasing the system size. We evaluate the performance of the repeated interaction model, specifically focusing on how often the mid-circuit measurements must be performed to recover the correct dynamics. In addition, we begin to validate fault-tolerant quantum algorithms based on repeated interactions by ensuring that generating initial states and factoring the overall dynamics into easily implementable, sequential operations through Trotter formulas is feasible. We perform this validation using numerical evaluations of the unitary dynamics generated by repeated interaction schemes, rather than by directly synthesizing and simulating quantum circuits. Our numerical simulations provide insight into the practical capabilities of quantum algorithms based on the repeated interaction model.

In Section 2, we outline the concepts from ET theory, which we use to construct our system Hamiltonian, as well as the Lindblad master equation used for the open quantum system description. We move on to explain how the non-unitary dynamics could be simulated on a unitary quantum computer through repeated interaction schemes, Trotter formulas, and state preparation. We then present the details of our numerical simulations in Section 3 and our results in Section 4. Finally, we provide a discussion of our results in Section 5 which ends with a summary of our main conclusions.

\section{Theory}
\subsection{Electron transfer dynamics}
For ET reactions, the key simulation target is the transfer rate of the electron moving from the donor to the acceptor. To simulate the transfer dynamics, rather than treating all the interactions between the electron and the nuclear degrees of freedom in the environment, we model the electron as interacting with a collective degree of freedom, called the reaction coordinate, which then interacts with the broader environment. The compression of the nuclear degrees of freedom can be visualized as replacing the nuclei with a single dipole that rotates to align with the electron, where the angle of the dipole is the reaction coordinate, as depicted in Figure \ref{fig:reaction_coord_pes}a. Our strategy for obtaining the transfer rate is to write a model Hamiltonian that extends the electronic system to a supersystem that includes a reaction coordinate. This has been shown to capture non-Markovian effects for systems with strong system-bath coupling and for structured environments.\cite{anto-sztrikacs_capturing_2021} We then use the Lindblad formalism to describe the interaction between the reaction coordinate and the environment, using jump operators to govern these interactions. These jump operators form the basis of the repeated interaction algorithm, which we use to simulate the open system dynamics over time.

\begin{figure}[ht]
    \centering
    \includegraphics{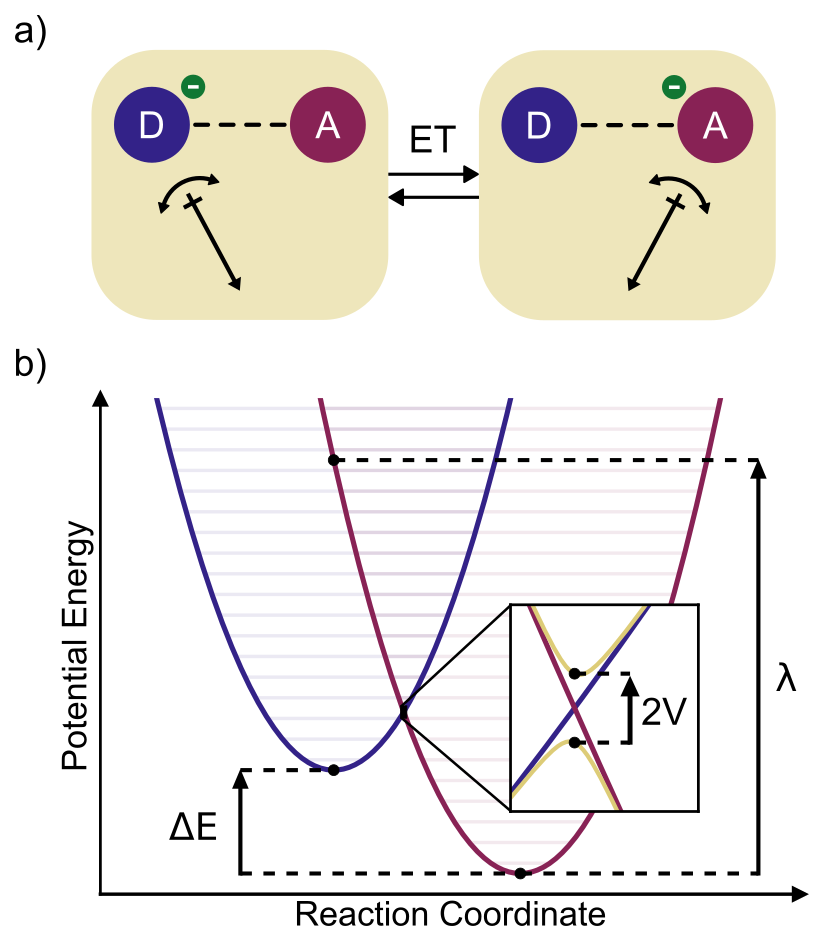}
        \caption{\textbf{Reaction Coordinate Model for Electron Transfer.} (a) Schematic representation of the reaction coordinate responding to the transferring electron. The nuclear degrees of freedom are approximated as a single dipole that aligns with the localization of the electron. In this picture, the angle of the dipole is the reaction coordinate. The dipole will rotate slightly around the equilibrium position due to thermal fluctuations. (b) Diabatic potential energy surfaces for the donor (blue) and acceptor (red) states as a function of the reaction coordinate. The inset shows the avoided crossing, with the adiabatic surfaces as yellow lines. Horizontal lines in the parabolas indicate quantized vibrational energy levels for each surface. Arrows show the donor-acceptor energy gap $\Delta E$, the reorganization energy $\lambda$, and two times the electronic coupling $2V$. Together, these visualizations illustrate how the reaction coordinate mediates electron transfer and how key energetic parameters shape the transfer dynamics.}
        \label{fig:reaction_coord_pes}
\end{figure}

The supersystem model for a donor-acceptor ET system has a main electronic system and a reaction coordinate. The electronic system is a two-level system with diabatic states $\ket{D}$ and $\ket{A}$ that represents the electron localized on either the donor or acceptor. We model the reaction coordinate as a single harmonic oscillator, where the equilibrium energy and position depend on the electron. The donor-acceptor Hamiltonian can be written using creation and annihilation operators $\hat{a}^\dagger$ and $\hat{a}$ for the reaction coordinate and Pauli matrix notation for the electronic system (i.e., $\sigma^{(z)} = \ket{D}\!\!\bra{D} - \ket{A}\!\!\bra{A}$ and $\sigma^{(x)} = \ket{D}\!\!\bra{A} + \ket{A}\!\!\bra{D}$) as
\begin{align}
    H_{\mathrm{DA}} = \frac{1}{2} \Delta E \sigma^{(z)} + V \sigma^{(x)} + \hbar \omega \hat{a}^{\dagger} \hat{a} + \sqrt{\hbar \omega \lambda} \sigma^{(z)} \hat{q}.
    \label{eq:H_DA}
\end{align}
The first term in this Hamiltonian gives the energy separation, $\Delta E$, between the donor and acceptor states, following the sign convention that $\Delta E$ is the donor energy minus the acceptor energy. The second term in the Hamiltonian is the electronic coupling, $V$, between the donor and acceptor states. The third term is the reaction coordinate harmonic oscillator with frequency $\omega$. The last term gives the coupling between the electron and the reaction coordinate. This coupling is between $\sigma^{(z)}$ for the electron and the position of the harmonic oscillator, $\hat{q} = \left(\hat{a}^\dagger + \hat{a}\right) / 2$, so it can be understood as shifting the potential energy minimum of the harmonic oscillator by $-\sqrt{\lambda/(\hbar \omega)}/2$ for the donor state and by $+\sqrt{\lambda/(\hbar \omega)}/2$ for the acceptor state. These shifts, along with the donor-acceptor energy gap, give rise to the two parabolas in the diabatic potential energy surface shown in Figure \ref{fig:reaction_coord_pes}b. The strength of this coupling is quantified by the reorganization energy, $\lambda$. The reorganization energy is the energy that would be released as the reaction coordinate goes from the minimum energy position of the acceptor state to the minimum energy position of the donor state while the electron is at the acceptor state. The adiabatic potential energy surface, which includes the effect of the electronic coupling, $V$, has an avoided crossing as shown in the inset of Figure \ref{fig:reaction_coord_pes}b.

The parameters in the Hamiltonian can take on a large range of chemically relevant values. When the electronic coupling, $V$, is small compared to the other energy scales, the reaction falls within the nonadiabatic regime.\cite{may_kuhn_2011} For nonadiabatic ET reactions, if the thermal energy is large relative to the oscillator energy ($k_\mathrm{B}T \gg \hbar\omega$), the rate of the reaction can be predicted with Marcus theory. Alternatively, if the thermal energy is comparable to or smaller than the oscillator energy, the reaction rate can be modeled with Fermi's golden rule. In the adiabatic regime, when the electronic coupling is comparable to other energy scales, the reaction rate cannot be modeled by Fermi's golden rule or Marcus theory. 

The DA model can be expanded to include bridging units, modeled as other diabatic states between the donor and acceptor. For the DBA system, the bridge states are labeled as $\ket{B_i}$ for the $i$th bridge unit from the donor. In this study, we included two bridging units in our version of the DBA model because the four total sites can be naturally encoded into two qubits. The Hamiltonian for a DBA system with two bridging units can be written as
\begin{align}
    H_{\mathrm{DBA}} =
    &\sum_\phi\left(\epsilon_\phi + \hbar \omega\left(\hat{q}-q_\phi\right)^2\right) \ket{\phi}\!\!\bra{\phi} + \hbar \omega \hat{p}^2 \notag\\
    &+ V \left( \ket{D}\!\!\bra{B_1} + \ket{B_1}\!\!\bra{B_2} + \ket{B_2}\!\!\bra{A} + \mathrm{h.c.} \right)
    ,
    \label{eq:H_DBA}
\end{align}
where the sum goes over $\phi = \{D, B_1, B_2, A\}$ and $\mathrm{h.c.}$ stands for Hermitian conjugate. The energy and equilibrium position of the reaction coordinate are given by $\epsilon_\phi$ and $q_\phi$ for each of the electronic diabatic states. This model assumes equal coupling between neighboring units and no long-range coupling between next-nearest neighbors or directly between the donor and acceptor.  We have also introduced the momentum of the reaction coordinate, $\hat{p}=i\left(\hat{a}^\dagger - \hat{a}\right)/2$. The Hamiltonian gives a diabatic potential energy surface that has four parabolas along the reaction coordinate, rather than the two for the DA model.

The unitary dynamics generated by the DA Hamiltonian (Equation \ref{eq:H_DA}) and the DBA Hamiltonian (Equation \ref{eq:H_DBA}) do not fully describe the reaction dynamics. Augmenting the electronic system with the reaction coordinate is only the first step in accounting for the effect of the nuclear degrees of freedom, and we must also consider the relaxation of the reaction coordinate. We use the Lindblad equation,
\begin{align}
    \frac{\partial \rho}{\partial t} = \mathcal{L} \rho= - \frac{i}{\hbar} \left[ H_\mathrm{ET}, \rho \right] + \mathcal{D}\left( \rho \right),
    \label{eq:Lindblad}
\end{align}
where $H_\mathrm{ET}$ could be either $H_{\mathrm{DA}}$ or $H_{\mathrm{DBA}}$, to describe the time dependence of the density matrix of the supersystem, $\rho$. The dissipator, $\mathcal{D}(\rho)$, has the form of a standard dissipator for the damped harmonic oscillator,
\begin{align}
    \mathcal{D}\left( \rho \right) = &\gamma \left( 1 + \bar{n} \right) \left( L \rho L^\dagger - \frac{1}{2} \left\{ L^\dagger L, \rho \right\} \right) \notag\\
    &+ \gamma \bar{n} \left(  L^\dagger \rho L - \frac{1}{2} \left\{ L L^\dagger, \rho \right\} \right),
    \label{eq:Dissipator}
\end{align}
with a decay rate of $\gamma$ and an average harmonic oscillator mode population given by a Boltzmann distribution, $\bar{n} = (\exp(\hbar \omega / (k_\mathrm{B}T)) - 1)^{-1}$, and where $\{A,B\}$ is the anticommutator given by $AB + BA$.\cite{breuer_theory_2002} The jump operators $L$ and $L^\dagger$ are shifted to account for the shift in the minimum energy position induced by the coupling between the electron and the reaction coordinate:
\begin{align}
    L = \hat{a} - \delta_\mathrm{ET},
    \label{eq:Jump_operator}
\end{align}
with the shift for the DA system given by
\begin{align}
    \delta_{\mathrm{DA}} = -\frac{1}{2} \sqrt{\frac{\lambda}{\hbar \omega}} \sigma^{(z)},
    \label{eq:shift_et}
\end{align}
and the shift for the DBA system given by
\begin{align}
    \delta_{\mathrm{DBA}} = \sum_\phi q_\phi\ket{\phi}\!\!\bra{\phi}.
    \label{eq:shift_dba}
\end{align}
By including this shift, the jump operators do not act directly on the local basis of the uncoupled reaction coordinate, which can prevent errors that come from working with local master equations.\cite{cattaneo_local_2019} A similar shift of the Lindblad jump operators was used recently to model proton transfer.\cite{zhang_modeling_2023}

We use the Lindblad equation to model the Markovian dynamics of the supersystem, which can, in turn, give rise to non-Markovian dynamics in the electronic system. The Lindblad equation relies on several assumptions: the separability of the initial state between the bath and system, and the Born-Markov and secular approximations.\cite{breuer_theory_2002} These assumptions are justified if there is weak coupling between the system and the bath. For modeling the ET supersystem dynamics with the Lindblad equation, we only assume that there is weak coupling between the reaction coordinate and the broader environment. We do not consider the case where the broader environment directly interacts with the electronic system. Previous studies have shown that such a setup preserves the non-Markovian dynamics of the main system for problems with a similar structure.\cite{anto-sztrikacs_capturing_2021} 

To quantify the degree of non-Markovianity retained in the electronic subsystem, we use a method introduced by Rivas, Huelga, and Plenio\cite{rivas_entanglement_2010} that we have labeled the RHP measure. Given that Markovian processes can only decrease entanglement, this approach evaluates how much entanglement is temporarily restored during the evolution of the system. Specifically, we simulate the dynamics of the electronic system maximally entangled with a non-interacting ancilla (separate from the ancilla used in the repeated interaction model). We then compute the concurrence, a measure of entanglement, between the electronic system and this non-interacting ancilla over time.

For purely Markovian dynamics, this entanglement will decrease monotonically due to the unidirectional flow of information from system to environment. However, a temporary increase in concurrence indicates a backflow of information, which is a hallmark of non-Markovian behavior. The RHP measure is found by integrating over the concurrence as
\begin{equation}
    \mathcal{I}^{(E)} = \int_{t_0}^{t_\mathrm{max}} | \frac{\mathrm{d}\mathcal{E}[\rho_\mathrm{SA}(t)]}{\mathrm{d}t} | \mathrm{d} t - \Delta \mathcal{E},
\end{equation}
with
\begin{equation}
    \Delta \mathcal{E} = \mathcal{E}[\rho_{SA}(t_0)] -\mathcal{E}[\rho_{SA}(t_\mathrm{max})]
\end{equation}
where $\mathcal{E}[\rho_\mathrm{SA}(t)]$ is the concurrence between the electronic system and the non-interacting ancilla at time $t$. A positive RHP measure indicates that the dynamics of the electronic system are non-Markovian. The RHP measure is a sufficient, but not necessary, condition to quantify non-Markovianity, meaning there could be a non-Markovian process that gives an RHP measure of zero.

\subsection{Non-unitary dynamics on a unitary quantum computer}
Simulating non-unitary dynamics on a unitary quantum computer presents distinct challenges. To perform a digital quantum simulation of an ET reaction, the non-unitary dynamics described by the Lindblad equation (Equation \ref{eq:Lindblad}) must be encoded in the operations available to a digital quantum computer---namely, unitary gates and measurement. In this study, we use a repeated interaction scheme, sometimes called a quantum collision model, to simulate non-unitary ET dynamics. The main idea of this scheme is to repeatedly perform a unitary evolution on the system of interest coupled to an ancilla system and then measure and reset the ancilla for each short time step. Figure \ref{fig:quantum_circuit} shows a quantum circuit diagram for this general scheme. Repeated interaction models are often studied in contexts where the discrete dynamics arise naturally, such as a system that occasionally collides with another system. However, this is not the case in this study. We are using a repeated interaction model because it can approximate Lindblad dynamics, even though repeated interactions do not obviously arise in ET theory.

\begin{figure}[ht]
    \centering
    \includegraphics{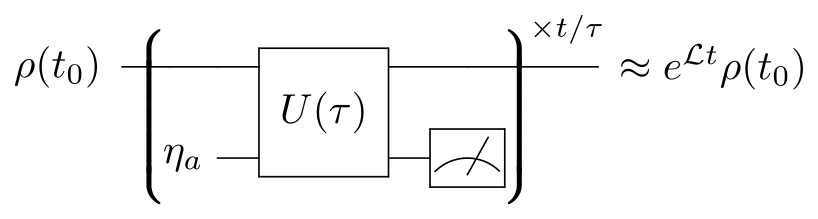}
        \caption{\textbf{General Quantum Circuit for the Repeated Interaction Model.} The first qubit register (top line) represents the system of interest, while the second qubit register (bottom line) represents the ancilla. $\rho(t_0)$ is the initial state of the system of interest and $\eta_a$ is the initial state of the ancilla qubit register. The system and ancilla have a joint unitary operation $U(\tau)$ that evolves the combined system for a short time $\tau$. The ancilla is then measured and reset to its initial state. The process in brackets is repeated $t/\tau$ times, allowing the circuit to approximate the Lindblad evolution of the system of interest over a total time $t$. This iterative approach enables the simulation of open quantum system dynamics within a fault-tolerant quantum computing framework.}
        \label{fig:quantum_circuit}
\end{figure}

We can precisely define the repeated interaction model as repeatedly applying a dynamical map to the system. The state of a system after $s$ interactions, each of duration $\tau$, is represented mathematically by 
\begin{align}
    \rho \left( t_0 + s \tau \right) = \ & \mathcal{E}_s\left[ \ldots \mathcal{E}_2\left[ \mathcal{E}_1\left[ \rho \left( t_0 \right) \right] \right] \right] \nonumber \\
    = \ & \mathcal{E}^s \left[ \rho \left( t_0 \right) \right],
    \label{eq:Repeated_map_applications}
\end{align}
where $t_0$ is the initial time. The dynamical map,
\begin{align}
    \mathcal{E} \left[ \rho \right] = \mathrm{Tr}_a \left\{ U(\tau) \left( \rho \otimes \eta_a\right) U^\dagger(\tau) \right\},
    \label{eq:RI_map}
\end{align}
describes how the system and the ancilla in some initial state $\eta_a$ undergo joint unitary evolution defined by $U$, before the ancilla system is traced out.

Following the work of Pocrnic et al.,\cite{pocrnic_quantum_2024} the joint unitary for the system-ancilla evolution is
\begin{align}
    U(\tau) = \exp\left\{ -\frac{i \tau}{\hbar} \left( H_\mathrm{ET} + \frac{1}{\sqrt{\tau}} H_\mathrm{int} \right)\right\},
    \label{eq:RI_unitary}
\end{align}
where $H_\mathrm{ET}$ is either $H_{\mathrm{DA}}$ or $H_{\mathrm{DBA}}$. The factor of $1/\sqrt{\tau}$ is a standard feature of repeated interaction models that creates a diverging coupling strength to the ancilla and preserves the dynamics in the limit of $\tau$ approaching zero.\cite{ciccarello_quantum_2022} In general, the ancilla system could also have internal dynamics described by operators that only act on the ancilla system; however, this is not needed to reproduce Lindblad dynamics. The Hamiltonian describing the interaction between the system and the ancilla is
\begin{align}
    H_\mathrm{int} = \sqrt{\gamma \left( 2 \bar{n} + 1\right)} \left( L \otimes \sigma_a^{(+)} + L^\dagger \otimes \sigma_a^{(-)} \right),
    \label{eq:H_I}
\end{align}
where $L$ and $L^\dagger$ are the jump operators defined by Equation \ref{eq:Jump_operator}, and $\sigma_a^{(+)} = \ket{1}_a\!\bra{0}$ and $\sigma_a^{(-)} = \ket{0}_a\!\bra{1}$ are the two-level system creation and annihilation operators, with the subscript $a$ denoting that they act on the ancilla qubit. Before each interaction, the state of the ancilla is
\begin{align}
    \eta_a &= \frac{1}{2 \bar{n} + 1} \big( \bar{n} \ket{0}_a\!\bra{0} + \left( \bar{n} + 1 \right) \ket{1}_a\!\bra{1} \big).
    \label{eq:Ancilla_initial_state}
\end{align}
This ancilla state is a mixed state that is diagonal in the computational basis of the qubit. During an implementation of the circuit, an external source of classical randomness could be used to initialize the ancilla in either the $\ket{0}_a$ state with probability $\bar{n}/(2 \bar{n} + 1)$ or in the $\ket{1}_a$ state with probability $(\bar{n} + 1)/(2 \bar{n} + 1)$.

To simulate an ET process, the dynamics start with an initial state of the supersystem, $\rho(t_0)$, which has the electronic system localized on the donor and the reaction coordinate equilibrated relative to the localized electron. This is described by
\begin{align}
    \rho \left( t_0 \right) = \ket{D}\!\!\bra{D} \frac{e^{-\beta \bra{D} H_\mathrm{ET} \ket{D}}}{Z},
    \label{eq:System_initial_state}
\end{align}
with $\beta = 1/(k_\mathrm{B}T)$ and a normalization constant $Z = \mathrm{Tr}\left\{ e^{-\beta \bra{D} H_\mathrm{ET} \ket{D}} \right\}$, where $H_\mathrm{ET}$ is either $H_{\mathrm{DA}}$ or $H_{\mathrm{DBA}}$. The exponential gives the equilibrated state of the reaction coordinate, which is a thermal harmonic oscillator state that has been displaced to the potential energy minimum position of the donor state. The initial state of the supersystem is also a mixed state, but due to the displacement, it is not diagonal in a convenient basis, meaning that there is not a trivial way to use an external source of classical randomness to initialize the state.

The displaced thermal state for the reaction coordinate can be prepared on a quantum computer using the same repeated interaction scheme that generates the dynamics of interest. The state preparation begins with the electron in the donor state and the reaction coordinate in an easy-to-prepare state, such as the ground state. A modification of the repeated interaction unitary (Equation \ref{eq:RI_unitary}) can be used to transform the easy-to-prepare state into the displaced thermal state when evolved for a long enough time. The state preparation procedure should not allow the electron to transfer from the donor, so electronic coupling to the donor should not be included. The state preparation (SP) unitary is then
\begin{align}
    U_\mathrm{SP}(\tau) = \exp \left\{ -\frac{i \tau}{\hbar} \left( H^{(0)}_{\mathrm{ET}} + \frac{1}{\sqrt{\tau}} H_\mathrm{int} \right) \right\},
    \label{eq:RI_unitary_SP}
\end{align}
where $H^{(0)}_{\mathrm{ET}}$ is either $H_{\mathrm{DA}} - V \sigma^{(x)}$ or $H_{\mathrm{DBA}} - V \left( \ket{D}\!\!\bra{B_1} + \ket{B_1}\!\!\bra{D} \right)$. In our numerical simulations, we monitored the progress of this state preparation procedure using the fidelity, defined as
\begin{align}
    F \left( \rho, \sigma \right) = \mathrm{Tr} \left\{ \sqrt{\rho^{1/2} \sigma \rho^{1/2}} \right\},
\end{align}
between the time-evolved state and the exact initial state defined in Equation \ref{eq:System_initial_state}.

The repeated interaction unitary (Equation \ref{eq:RI_unitary}) cannot be implemented directly during a digital quantum simulation because the exponential contains terms that would need to act on the same qubits simultaneously. We therefore use a Trotter formula to split the unitary into factors that can each be implemented sequentially for a small time step.\cite{nielsen_quantum_2010} Implementing these factors one at a time introduces an error when the factors do not commute. This error can be suppressed by using higher-order Trotter formulas or decreasing the duration of the time step. For the repeated interaction scheme, the total time needed for the dynamics, $t$, is already split into steps of length $\tau$. With the Trotter formula, each $\tau$ step will be further factored into $n$ Trotter steps. Given some value of $\tau$ needed to reproduce Lindblad dynamics, the goal for implementing the Trotter formula is to optimize the number of Trotter steps per $\tau$ by balancing accuracy with the number of operations.

To numerically investigate this trade-off between accuracy and the number of operations for the donor-acceptor ET simulations, we use a 2nd-order Trotter formula,\cite{suzuki_fractal_1990} 
\begin{align}
    &U \left( \tau \right) \approx U_{\mathrm{Trotter}} \left( \tau \right) = \nonumber \\
    & \left( \prod_{k=1}^{6} \exp\left\{ -\frac{i}{\hbar} \frac{\tau}{2n} H_k \right\} \prod_{k=6}^{1} \exp\left\{ -\frac{i}{\hbar} \frac{\tau}{2n} H_k \right\} \right)^n .
    \label{eq:Trotter}
\end{align}
The six $H_k$ factors come from the four terms in $H_{\mathrm{DA}}$ and the two terms in $H_\mathrm{int} / \sqrt{\tau}$. These six factors are sorted in descending order so the $k=1$ factor has the largest coefficient. The second product in this Trotter formula reapplies the $H_k$ factors in the opposite order, which is the key difference between 1st- and 2nd-order Trotter formulas. In a hardware-specific implementation, the six factors might be further divided based on which specific gates are available and how the operators are encoded, but this is not considered in this study.

\section{Numerical Simulation Details}
To study how a repeated interaction scheme can be used in a quantum algorithm to model ET, we performed numerical simulations using the four parameter sets listed in Table \ref{tab:param}. We chose these parameter sets to systematically evaluate how the repeated interaction algorithm scales while ensuring the numerical simulations were feasible, rather than attempting to model specific systems. This means that the parameter sets do not unambiguously fall into single, well-defined chemical regimes, such as nonadiabatic vs. adiabatic. The first parameter set, which we label as weakly coupled because of the small electronic coupling, $V$, serves as the reference for the other three sets. The strongly damped parameter set increases the damping rate, $\gamma$, of the reaction coordinate, representing a system with stronger dissipation. The strongly coupled parameter set features a larger electronic coupling while keeping all other parameters identical to the weakly coupled parameter set. Finally, the high-temperature parameter set increases the thermal energy, $k_\mathrm{B}T$, along with the accompanying higher reorganization energy, $\lambda$, reflecting conditions where classical transfer effects become more prominent. The three parameter sets with $k_\mathrm{B}T=1$ are collectively labeled as low temperature.

In addition to the four DA parameter sets, we also investigated a DBA system. This system was configured with the same electronic coupling, thermal energy, and damping rate as the weakly coupled parameter set. The spacing between all neighboring units was set to $\hat{q} = 1$, matching the DA distance in the weakly coupled case. Relative to the acceptor energy, the site energies of the donor and bridge units were chosen as $\epsilon_D=5 \hbar\omega$, $\epsilon_{B_1}=4 \hbar\omega$, and $\epsilon_{B_2}=3 \hbar\omega$.

\begin{table*}[ht]
    \caption{\textbf{Simulation Parameters for Donor-Acceptor Electron Transfer.} The weakly coupled parameter set serves as the reference for the other three parameter sets. Each of the remaining sets varies a single parameter to explore different electron transfer regimes, except for the high-temperature set, where both the thermal energy and the reorganization energy are increased due to their inherent dependence on each other.}
    \begin{tabular}{@{}l p{2cm} p{2.5cm} p{2cm} p{2.5cm}@{}} \hline
        Parameter set & Electronic \newline coupling, $V \, (\hbar \omega)$ & Reorganization energy, \newline $\lambda \, (\hbar \omega)$ & Thermal \newline energy, $k_\mathrm{B}T \, (\hbar \omega)$ & Damping \newline rate, \newline $\gamma \, (\omega / 2 \pi)$ \\
        \hline
        Weakly coupled   & 0.1 & 1  & 1 & 0.01 \\
        Strongly damped  & 0.1 & 1  & 1 & $1/(2\pi) \approx 0.16$ \\
        Strongly coupled & 1   & 1  & 1 & 0.01 \\
        High temperature & 0.1 & 20 & 2 & 0.1  \\
        \hline
    \end{tabular}\\
    \label{tab:param}
    \vspace{2pt}
\end{table*}

We truncated the Hilbert space of the reaction coordinate in our numerical simulations, which would also be required for the simulation of a harmonic oscillator on a digital quantum computer. For the low-temperature parameter sets, the reaction coordinate was truncated to 16 energy levels, corresponding to a four-qubit binary encoding, while for the high-temperature parameter set, it was truncated to 32 energy levels, requiring five qubits. We ensured the truncation did not appreciably affect the dynamics by inspecting the expectation value of the initial position for the harmonic oscillator (see Figure \ref*{fig:pos} in the Supplementary Material) and the transfer rate error from Lindblad simulations for different numbers of qubits (Figures \ref*{fig:trunc}–\ref*{fig:trunc3}). It is interesting to note that rather than thermal energy, the primary factor determining the necessary truncation was the displacement of the minimum energy positions, given by $\pm \sqrt{\lambda/(\hbar \omega)}/2$ for the DA systems. This is because displaced oscillator states require a large number of basis states to be accurately represented in the original oscillator basis, even when thermal energy is relatively small.

We used QuTiP version 5.0.2 to perform the numerical integration of the Lindblad master equation and the repeated interaction scheme. Each simulation was run over a time interval from $t=0$ to $t=1000 (2\pi / \omega)$. To calculate the transfer rate $k$, the donor population was fit using SciPy version 1.13.1 to an exponential decay of the form $P_D(t) = P_D(0)\exp{-kt}$ where $P_D$ is the donor population.

\section{Results}
We show the donor populations as a function of time in Figure \ref{fig:pop} from repeated interaction simulations (numerical evaluations of Equation \ref{eq:Repeated_map_applications}) for the four DA parameter sets---weakly coupled, strongly damped, strongly coupled, and high temperature---with values of the energy gap, $\Delta E$, that led to fast transfer. The repeated interaction model successfully reproduced the Lindblad dynamics (numerical evaluations of Equation \ref{eq:Lindblad}) using an interaction duration of $\tau = 0.1 (2\pi/\omega)$ for the low-temperature parameter sets and $0.01 (2\pi/\omega)$ for the high-temperature parameter set. We saw coherent oscillations in the donor population from the Lindblad and converged repeated interaction simulations. In general, when the repeated interaction duration was too large, we saw too many oscillations, resembling unitary dynamics.

\begin{figure*}[ht]
    \centering
    \includegraphics{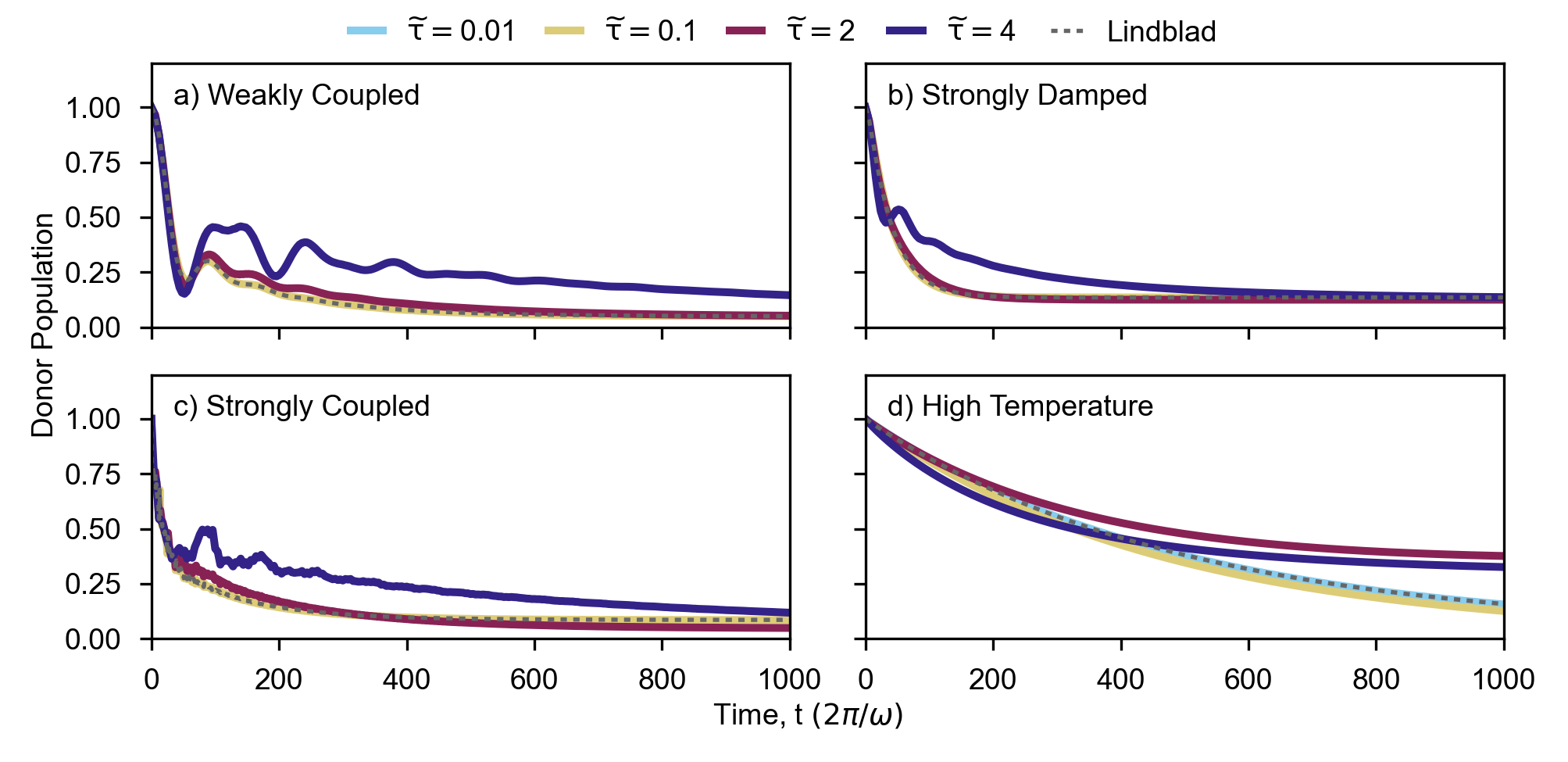}
        \caption{\textbf{Repeated Interaction Simulations.} Donor populations from repeated interaction simulations of the (a) weakly coupled, (b) strongly damped, (c) strongly coupled, and (d) high-temperature parameter sets as a function of time, $t$. Colors show the donor populations for different values of the repeated interaction duration, with $\widetilde{\tau}=\tau\omega/(2\pi)$. Dashed lines show reference populations from Lindblad simulations. The energy separations $\Delta E$ were set to: $3\hbar \omega$ for weakly coupled, $2\hbar\omega$ for strongly damped, $4.4\hbar \omega$ for strongly coupled, and $20\hbar \omega$ for high-temperature parameter sets. The agreement between the repeated interaction and Lindblad simulations across all parameter sets demonstrates the reliability of the repeated interaction model in capturing electron transfer dynamics.}
        \label{fig:pop}
\end{figure*}

In Figure \ref{fig:dba}, we show the donor population from repeated interaction simulations with different interaction durations, along with the population dynamics of each site of the DBA model. Similar to the low-temperature DA parameter sets, the DBA simulation converged to the Lindblad result with an interaction duration of 0.1 $(2\pi/\omega)$. The DBA model exhibited a quicker transfer from the donor state compared to the DA simulations, suggesting that the inclusion of bridges provides either an effectively stronger coupling or more favorable dynamics. The DBA system exhibited hopping behavior, characterized by the sequential transfer of population between localized states as seen in Figure \ref{fig:dba}b. Initially, the population resided on the donor, which then transferred to the first bridge site, followed by the second bridge site, and then ultimately to the acceptor. This stepwise transfer reflects the intermediate role of the bridge sites in facilitating the ET process. The dynamics of hopping are influenced by the coupling strengths between adjacent sites, as well as the interaction with the environment, which can introduce decoherence or dissipation. Furthermore, coherent oscillations were observed between the states, which were evident in all low-temperature parameter sets explored in this study. In more classical or strongly damped conditions, the transfer should appear smoother and less oscillatory.

\begin{figure*}[ht]
    \centering
    \includegraphics{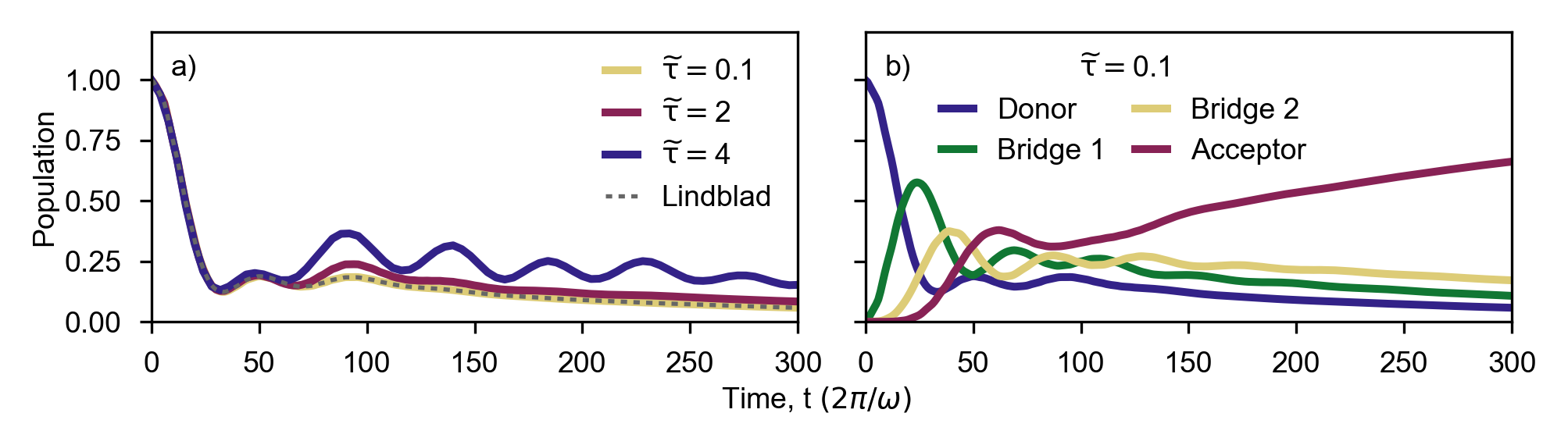}
        \caption{\textbf{Donor-Bridge-Acceptor Populations.} 
        Colors in (a) show the donor population for different values of the repeated interaction duration, with $\widetilde{\tau}=\tau\omega/(2\pi)$ for the donor-bridge-acceptor parameter set. The dashed line shows the reference donor population from a Lindblad simulation.
        (b) Populations of the donor, bridges, and acceptor from repeated interaction simulations with $\tau=0.1(2\pi/\omega)$. The sequential population transfer from the donor to the acceptor via the bridge sites reflects a stepwise hopping mechanism with coherent oscillations. These results illustrate how the repeated interaction model captures intermediate bridge site dynamics, highlighting its ability to resolve multi-step electron transfer processes, as well as investigating how the algorithm scales when the system size is increased.}
        \label{fig:dba}
\end{figure*}

The transfer dynamics simulated with the repeated interaction model also converged across values for the donor-acceptor energy gap $\Delta E$ to the Lindblad results for the low-temperature DA systems with a repeated interaction duration of $0.1 (2\pi/\omega)$. In Figure \ref{fig:ri_rates}, the transfer rates obtained from repeated interaction simulations with different values of $\tau$ are plotted against the DA energy gap. Figure \ref{fig:ri_rates}a shows the transfer rates for the weakly coupled parameters, where narrow peaks are seen on resonance with integer multiples of $\hbar \omega$. Figure \ref{fig:ri_rates}b is the transfer rates for the strongly damped parameter set, where as expected the peaks are broader than the peaks from the weakly coupled parameter set. Figure \ref{fig:ri_rates}c shows the rates for the strongly coupled parameter set. With these parameters, we see peaks that are even broader and no longer on resonance with $\hbar \omega$. For both the weakly coupled and the strongly coupled parameters, using too large of an interaction duration led to peaks at the same value of $\Delta E$ but with a smaller transfer rate than the Lindblad results. Interestingly for the strongly damped case, using too large of an interaction duration led to peaks that are larger or smaller than the Lindblad result, and that are not centered on the same value of $\Delta E$.

\begin{figure*}[ht]
    \centering
    \includegraphics{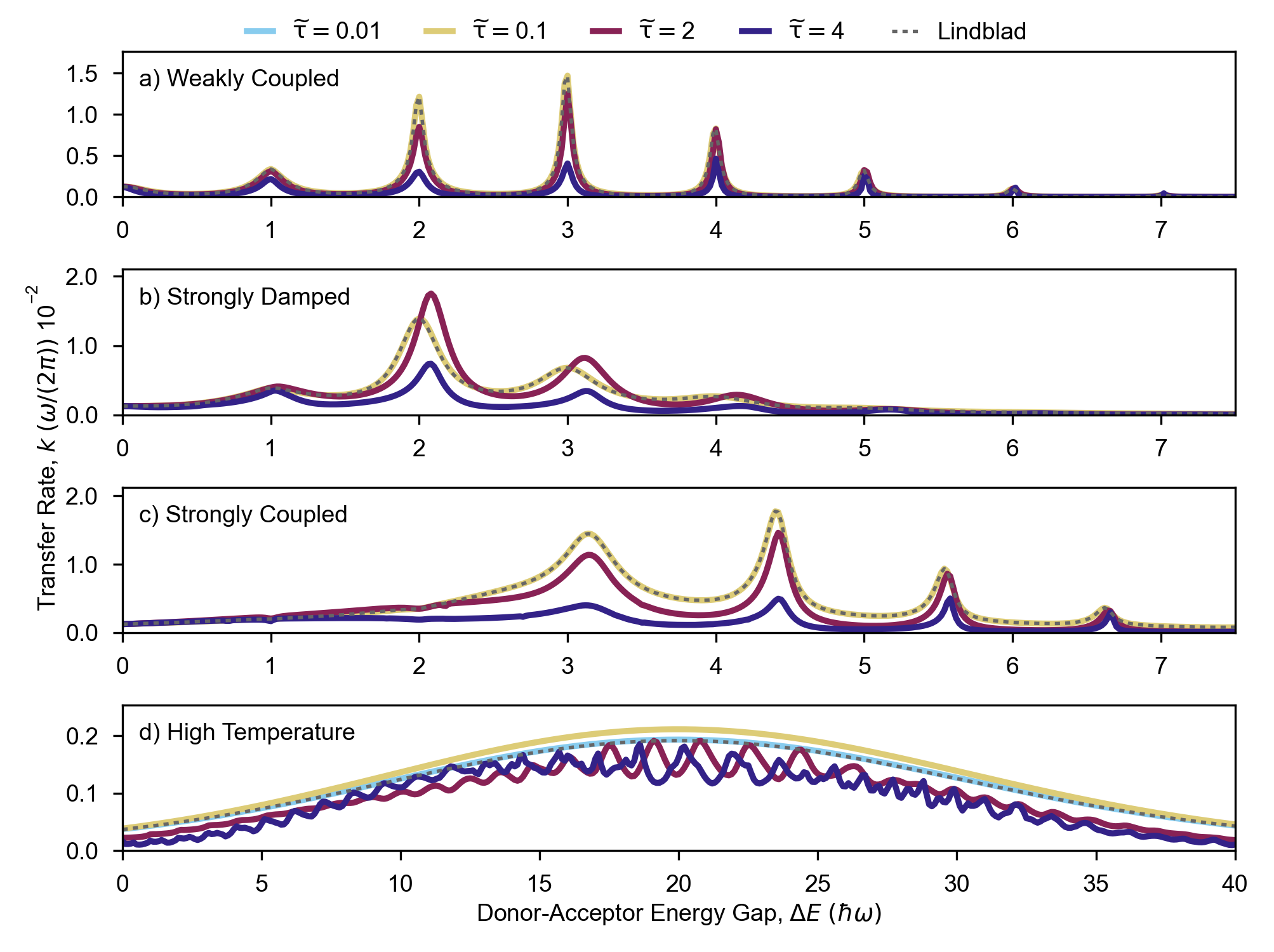}
        \caption{\textbf{Electron Transfer Rates.} Transfer rates, $k$, from repeated interaction simulations of the weakly coupled (a), strongly damped (b), strongly coupled (c), and high-temperature (d) parameters as a function of donor-acceptor energy separation, $\Delta E$. Colors show the transfer rates for different values of the repeated interaction duration, with $\widetilde{\tau} = \tau \omega / (2 \pi)$. Dashed lines show reference transfer rates from Lindblad simulations. The consistency between the repeated interaction and Lindblad simulations across all parameter sets validates the accuracy of the repeated interaction model in capturing electron transfer dynamics.}
        \label{fig:ri_rates}
\end{figure*}

Figure \ref{fig:ri_rates}d shows the transfer rates for the high-temperature parameter set. For the Lindblad simulation, the transfer rate initially increased as the energy gap increased but then decreased after the energy gap was larger than the reorganization energy of 20 $\hbar \omega$, showing Marcus inverted region behavior. Repeated interaction simulations with larger interaction durations showed this same general behavior but had numerous small oscillations, rather than a single broad peak. As the repeated interaction duration decreased, the oscillations in the transfer rate disappeared. For $\tau = 0.1 (2\pi/\omega)$, the transfer rates were slightly larger than the Lindblad results, but with $\tau = 0.01 (2\pi/\omega)$, the results converged to the Lindblad rates, meaning that ten times as small of a step was needed compared to the low-temperature DA parameters.

We evaluated the non-Markovian character of the electronic system from Lindblad simulations by calculating the RHP measure as a function of the donor–acceptor energy gap, as shown in Figure \ref{fig:rhp_lindblad}. The three low-temperature parameter sets all showed non-Markovianity, which is evident from their non-zero RHP measures. The RHP measure for the high-temperature parameter set was equal to zero for all values of the donor-acceptor energy gap. For the weakly coupled and strongly coupled systems, we observed distinct dips in the RHP measure at the donor-acceptor energy gaps associated with high transfer rates. This is likely because higher transfer rates allow the reaction coordinate to thermalize faster, and after the reaction coordinate thermalizes, the backflow of information stops.
In contrast, the strongly damped regime exhibited consistently low RHP values that are still greater than zero, but about an order of magnitude smaller than the values for the weakly coupled regime. This indicates that strong system-bath interactions suppress memory effects and push the system closer to the Markovian regime.

To confirm that the repeated interaction model preserves both the non-Markovian character of the dynamics and the RHP measure itself, we performed a simulation of the repeated interaction model with the ancilla for the RHP measure for the weakly coupled parameter set with $\Delta E = 3\hbar \omega$. The results, shown in the Supplementary Material (Figure \ref*{fig:rhp_ri}), are consistent with those from the Lindblad simulation. Specifically, we observed the same increases in entanglement, confirming that as the repeated interaction model converges to the Lindblad limit, it preserves the backflow of information from the reaction coordinate to the electronic system.

\begin{figure}[ht]
    \centering
    \includegraphics{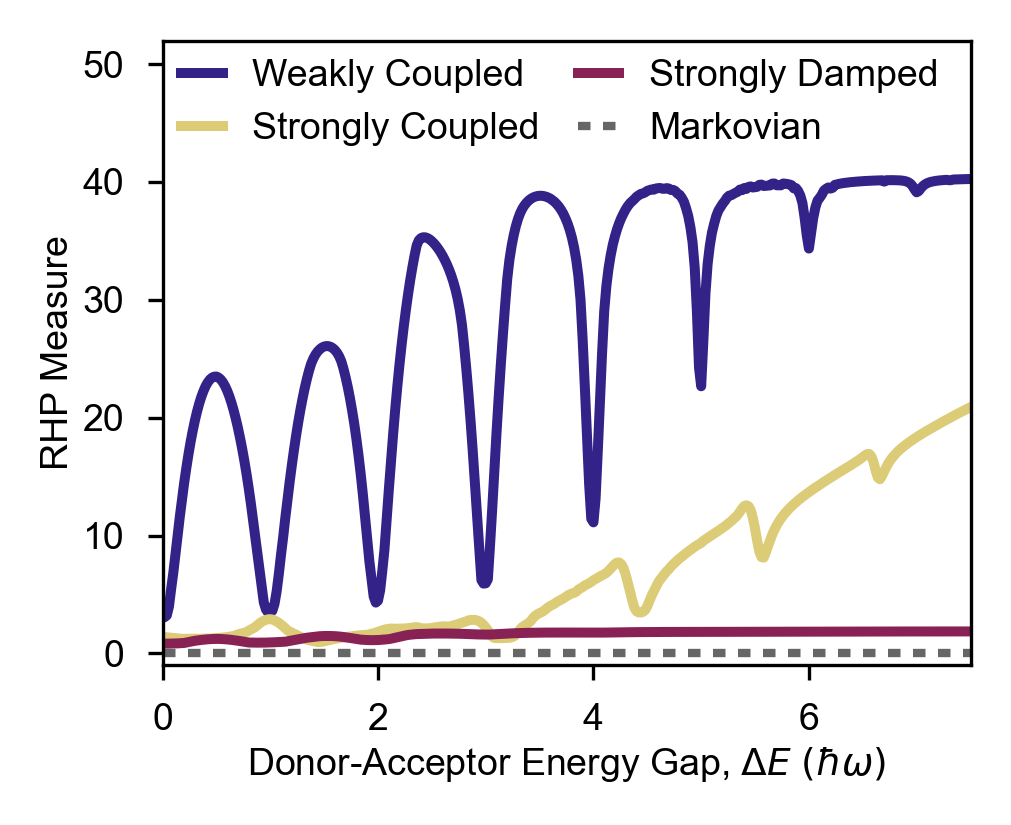}
        \caption{\textbf{Rivas, Huelga, Plenio Measures for the Donor-Acceptor Parameter Sets.} The measure is non-zero, indicating non-Markovian dynamics, for all values of the donor-acceptor energy gap for the low-temperature parameter sets.}
        \label{fig:rhp_lindblad}
\end{figure}

By evolving an easy-to-prepare state with a repeated interaction model that did not contain the electronic coupling, we prepared states that had high fidelity with the exact initial state (Equation \ref{eq:System_initial_state}) for all parameter regimes for the DA model and the DBA system. As shown in Figure \ref{fig:state_prep}, the state preparation procedure generated states with above 99\% fidelity compared to the exact initial states within $t=400 (2\pi/\omega)$ for all the parameter sets. Note that because the exact initial state does not depend on the electronic coupling or the damping rate, the three low-temperature ($k_\mathrm{B}T=1\hbar\omega$) parameter sets have the same initial state. The fidelity for the high-temperature parameters rose quickly and then plateaued, while for the weakly coupled and DBA systems, it increased slowly. Even though the parameter sets led to different behaviors in how the fidelity increases, the state preparation procedure was still capable of generating high-fidelity states after a short time evolution.

\begin{figure}[ht]
    \centering
    \includegraphics{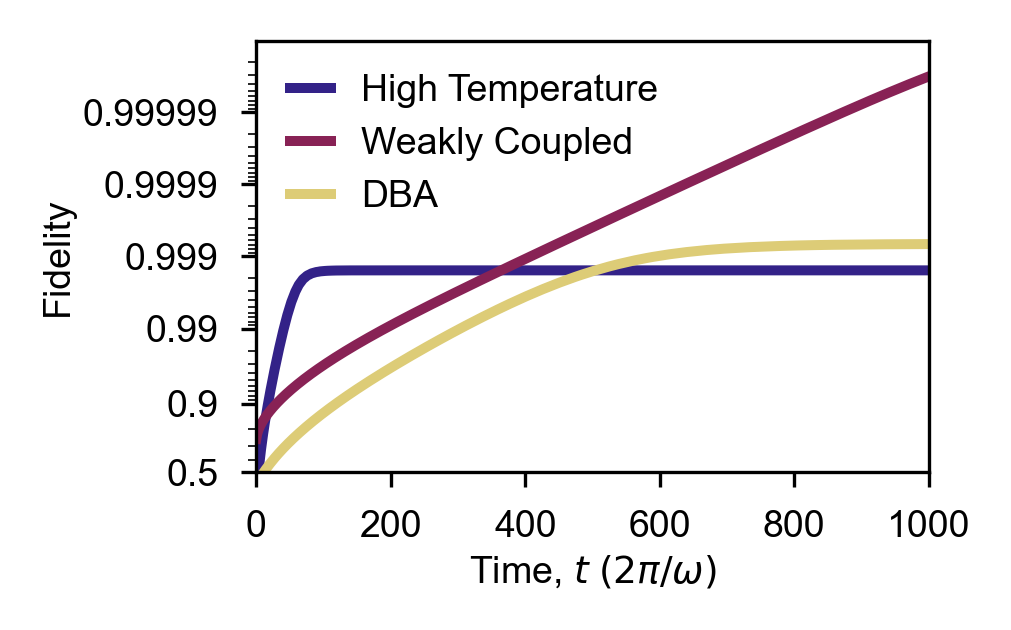}
        \caption{\textbf{Fidelity of Prepared Initial State.} Fidelity between the exact initial state and the state following the state preparation procedure as a function of simulation time for the weakly coupled, high-temperature, and donor-bridge-acceptor (DBA) parameter sets. The state preparation begins with the electronic system in the donor state and the oscillator in the ground state and then evolves in time with a repeated interaction model that does not contain electronic coupling. The three parameter sets with $k_\mathrm{B}T=1\hbar\omega$ have the same initial state, so we only show the weakly coupled set. The fidelity quickly increases across all parameter sets, confirming that the state preparation scheme reliably initializes the system for repeated interaction simulations.}
        \label{fig:state_prep}
\end{figure}

Factoring the repeated interaction unitary using a 2nd-order Trotter formula (Equation \ref{eq:Trotter}) led to small percent errors in the calculated transfer rates for all of the parameter sets tested for the DA model system. For a single Trotter step per repeated interaction ($n=1$), the percent error in the transfer rate was less than 4\% for all of the parameter regimes, as shown in Figure \ref{fig:trotter}, where the energy gap, $\Delta E$, was chosen to give the largest transfer rate. The percent error was the smallest for the high-temperature parameters, likely due to the shorter interaction duration needed to converge to Lindblad dynamics. Because of the small percent error and the increased computational demand for simulations of the high-temperature parameters, we did not go beyond 25 Trotter steps per repeated interaction. Interestingly, for all the parameter sets, the percent error in the transfer rate did not monotonically decrease as the number of Trotter steps increased. For the low-temperature parameters, the percent error had a local minimum between $n=2$ and $n=5$, while for the high temperature, $n=1$ was a local minimum. Above 30 Trotter steps per repeated interaction, the percent error steadily decreased for the low-temperature sets.

\begin{figure*}[ht]
    \centering
    \includegraphics{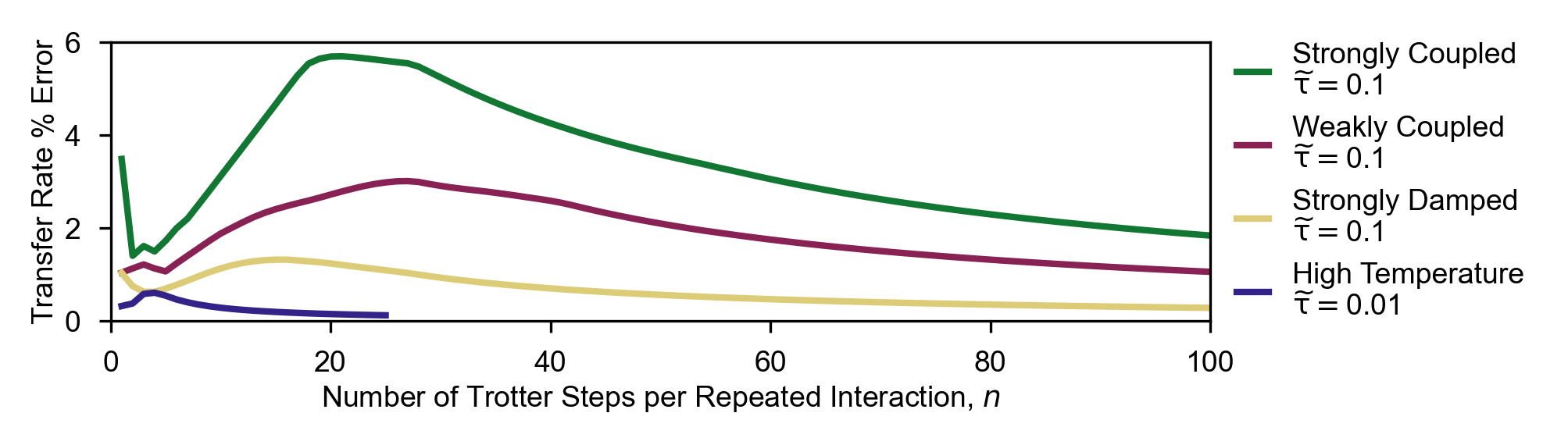}
        \caption{\textbf{Error from Using Trotter Formula.} Percent error in the transfer rates as a function of the number of Trotter steps per repeated interaction $n$ from 2nd-order Trotter simulations of repeated interaction schemes relative to exact simulations with no Trotterization. The repeated interaction duration, labeled with $\widetilde{\tau} = \tau \omega / (2 \pi)$, is ten times shorter for the high-temperature case. The energy separations $\Delta E$ were set to the largest peaks: $4.4\hbar \omega$ for strongly coupled, $3\hbar \omega$ for weakly coupled, $2\hbar\omega$ for strongly damped, and $20\hbar \omega$ for high-temperature parameter sets. These results demonstrate that Trotterization error decreases with increasing Trotter steps, with the strongest errors observed in strongly coupled systems which has faster dynamics. For high-temperature systems, the same would be expected, but the finer time discretization required for this parameter set alleviates the need for more Trotter steps per repeated interaction.}
        \label{fig:trotter}
\end{figure*}

\section{Discussion}
Our numerical simulations of the repeated interaction model successfully generated non-Markovian ET dynamics without needing prohibitively small repeated interaction duration for any of the low-temperature parameter regimes tested. Additionally, the model reproduced Marcus behavior for the high-temperature parameter set and showed a favorable scaling. We also confirmed that the non-Markovianity was preserved using the repeated interaction scheme. A limited discussion of the scaling of the model with respect to the analytical results of Pocrnic et al.\cite{pocrnic_quantum_2024} can be found in Section {\ref*{sec:scale}} in the Supplemental Material. These results show promise for applying quantum algorithms based on repeated interaction models to open chemical systems. 

State preparation was straightforward using the repeated interaction method, where input states with over 99\% fidelity were generated over a short time evolution. The practical utility of this method was further confirmed by examining how the error introduced by the Trotter formula scales with the number of Trotter steps. For the systems studied, we found that a single Trotter step per repeated interaction produced an acceptable error of less than 4\%. While the number of Trotter steps could be optimized for these types of systems, we expect adequately low Trotter errors to be found in just a few Trotter steps per repeated interaction. Seeing as the dynamics are already discretized into small steps for the repeated interaction method, it is unsurprising that a single Trotter step per repeated interaction was sufficient. Taken together, this seems like a promising avenue for simulating non-unitary dynamics for chemistry on a unitary quantum computer.

The SVD-based approach, as demonstrated in simulations of the FMO complex,\cite{hu_general_2022,dan_simulating_2025} effectively maps non-unitary dynamics onto quantum circuits, making it well-suited for NISQ hardware. However, its dependence on classical diagonalization introduces a scaling bottleneck, particularly for larger systems or those with strong correlations and memory effects. While quantum singular value transformation has been proposed as a potential way to bypass this classical step,\cite{suri_two-unitary_2023} the necessity of diagonalization remains a fundamental constraint when dealing with large systems.

By contrast, the repeated interaction model provides a direct encoding of the system-environment interaction into the quantum simulation, without the need for matrix decomposition. This makes it a promising alternative, especially for scenarios where environmental effects are significant and non-trivial to approximate classically. Given that we observe convergence to Lindblad dynamics with relatively few iterations, repeated interaction avoids the computational overhead that might otherwise be a concern.

Ultimately, the suitability of each method depends on the problem at hand. For relatively simple, well-characterized dynamics like those of the FMO complex, SVD remains a powerful tool. However, for more complex open quantum systems where system-environment interactions play a crucial role, repeated interaction, and other variational approaches may offer a more scalable and physically transparent framework. The trade-offs between classical preprocessing, circuit depth, and accuracy should be carefully considered when selecting the most appropriate algorithm.

Recent analog quantum simulations of ET reactions on real quantum hardware have demonstrated the potential of near-term quantum devices for studying open quantum systems. Trapped-ion and superconducting qubit platforms have been used to model dissipative and coherent dynamics, providing experimental validation of theoretical frameworks.\cite{schlawin_continuously_2021} However, despite these advances, current analog simulations remain restricted to parameter regimes set by the hardware, preventing the exploration of classical reaction limits. This highlights a fundamental limitation of analog approaches, as they cannot easily access the full range of system-bath interactions predicted in theoretical models.

Different quantum approaches will likely be suited to different aspects of open quantum system simulations. Digital algorithms such as repeated interactions, quantum singular value transformations, and variational master equation solvers provide a flexible framework that can be adapted to various system-environment interactions. To get a practical sense of how this repeated interaction scheme might run on current quantum hardware, you would need to choose an encoding scheme for the harmonic oscillator and then compile the algorithm to a quantum circuit. The measurement and resetting of the ancilla are likely to be slow and noisy steps in the execution of the circuit, so special attention would need to be given to the measurement rate and the mid-circuit measurement crosstalk error, which would likely make running the circuit unobtainable without error correction. Before error correction is available, analog quantum simulations may offer advantages for specific systems if the experimental system naturally mimics the desired interaction, has strong intersystem correlations compared to system-environment correlations, or has a structured connectivity which fits with a specific device.

As fault-tolerant quantum hardware becomes more accessible, the scalability and programmability of digital methods will make them increasingly dominant, particularly for simulating complex chemical dynamics with strong system-bath correlation and highly structured environments. Key developments to watch include advances in quantum error correction, the implementation of noise-resilient gates, and improved encodings of open system dynamics onto logical qubits. Ultimately, the choice of method will depend on the nature of the system under study, with digital and analog approaches serving complementary roles in different problem regimes.

\section{Conclusion}
We have shown that a repeated interaction model can reproduce non-Markovian ET dynamics across realistic chemical parameter regimes. Additionally, we have shown that the model scales well in terms of the number of repeated interactions needed as the temperature increases. When fault-tolerant quantum hardware becomes available, we expect quantum algorithms based on repeated interaction schemes to be useful for modeling real systems, where more complexity could be incorporated by adding structured baths, multidimensional reaction coordinates, or more intricate coupling schemes. Our findings suggest that repeated interaction models provide an efficient and scalable approach to simulating open quantum systems, particularly for parameter regimes relevant to ET. In future work, it would be useful to synthesize and simulate quantum circuits of repeated interaction algorithms, in order to directly study the scaling of the number of qubits and circuit depth needed to simulate chemical systems. Additionally, it would be valuable to benchmark this approach against both classical and quantum algorithms for larger and more complex molecular systems with applications such as enzymatic catalysis or molecular electronics. A key open question is whether scalable, repeated interaction algorithms can be designed to reproduce non-Markovian dynamics of chemical systems without needing to use the reaction coordinate supersystem approach. As quantum hardware advances, we hope quantum simulations based on repeated interactions will help chemists elucidate real-world chemical dynamics at the quantum level.

\section*{Supplementary Material}
    See the Supplementary Material for the harmonic oscillator truncation analysis, the evaluation of how the repeated interaction model preserves non-Markovianity, and a comparison of the scaling of the algorithm with an analytical bound.
    %

\begin{acknowledgments}
We thank Prof. Dvira Segal and Matthew Pocrnic for fruitful discussions regarding the repeated interaction model. This work is supported by the Novo Nordisk Foundation, grant number NNF22SA0081175, NNF Quantum Computing Programme. M.S.T.\ and G.C.S.\ acknowledge support from the Novo Nordisk Foundation, grant number NNF20OC0060019, Challenge Project ``Solid-Q''.
\end{acknowledgments}

\section*{Author Declarations}
\subsection*{Conflict of Interest}
The authors have no conflicts to disclose.

\subsection*{Author Contributions}
\textbf{Lea K. Northcote}:
Conceptualization (equal);
Data curation (equal);
Formal analysis (equal);
Investigation (equal);
Methodology (equal);
Software (equal);
Validation (equal);
Visualization (equal);
Writing – original draft (equal);
Writing – review \& editing (equal).
\textbf{Matthew S. Teynor}:
Conceptualization (equal);
Data curation (equal);
Formal analysis (equal);
Investigation (equal);
Methodology (equal);
Software (equal);
Validation (equal);
Visualization (equal);
Writing – original draft (equal);
Writing – review \& editing (equal).
\textbf{Gemma C. Solomon}:
Conceptualization (supporting);
Funding acquisition (lead);
Project administration (lead);
Resources (lead);
Supervision (lead);
Writing – review \& editing (supporting).

\section*{Data Availability}
The data that support the findings of this study are openly available in the Electronic Research Data Archive at \url{http://doi.org/10.17894/ucph.a852a077-7b9b-4f4c-9a6d-1a0f3c79bd2e}. The code used to produce the simulation data is available on GitHub at \url{https://github.com/mteynor/ri4et}.

\section*{References}
\bibliography{bib}

\clearpage


\section*{Supplementary material}

# Supplementary Material:
# Repeated Interaction Scheme for the Quantum Simulation of Non-Markovian Electron Transfer Dynamics

Lea K. Northcote,[1, *)] Matthew S. Teynor,[1, 2, *)] and Gemma C. Solomon[1, 2, †)]
[1)]*NNF Quantum Computing Programme, Niels Bohr Institute, University of Copenhagen, DK-2100 Copenhagen Ø, Denmark*
[2)]*Nano-Science Center and Department of Chemistry, University of Copenhagen, DK-2100 Copenhagen Ø, Denmark*

(Dated: 1 May 2025)

## I. HARMONIC OSCILLATOR TRUNCATION

For numerical simulations of harmonic oscillators, we have to truncate the infinite-dimensional Hilbert space to a finite number of energy levels. This truncation is also required for digital quantum simulations, where the oscillator is encoded into a qubit register. With binary encoding, $n$ qubits can encode $2^n$ energy levels. To systematically choose the number of qubits required to encode the harmonic oscillator in the electron transfer (ET) model systems, we evaluated the initial position as well as the transfer rate error from Lindblad simulations for different numbers of qubits. Figure S1 shows the position of the truncated harmonic oscillator for the initial states of the weakly coupled (which has the same initial state as the strongly coupled and strongly damped parameters) and high-temperature donor-acceptor systems, compared to the exact initial position $-\sqrt{\lambda/(\hbar\omega)}/2$. The weakly coupled case has the correct initial position when there are at least four qubits (16 energy levels). The high-temperature system requires five qubits (32 energy levels) before it has the correct initial position. Figure S2 and Figure S3 show the transfer rates for three, four, and five qubits, as well as the percent error in the transfer rate relative to using six qubits. We achieve a maximum error of less than 2% in the transfer rate with four qubits for the weakly coupled and five qubits for the high-temperature parameter sets.

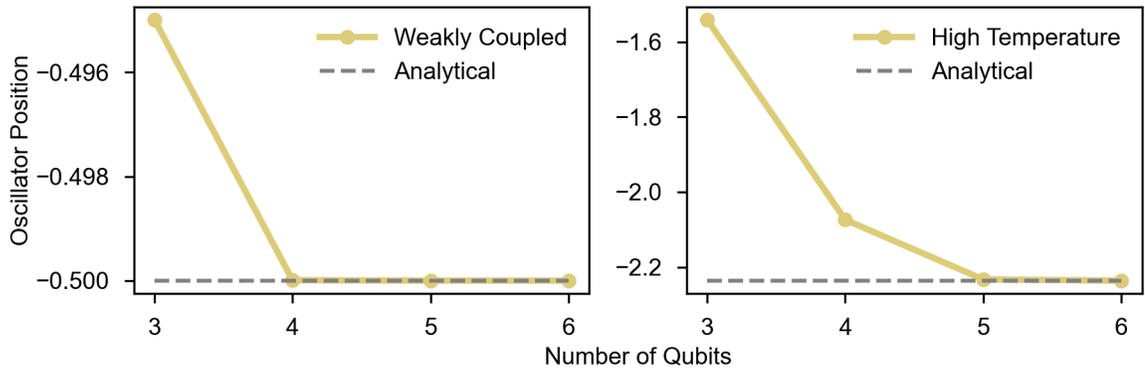

FIG. S1. **Expectation Value of Position Operator for the Harmonic Oscillator Across Regimes.** The first oscillator position in the simulations was evaluated for $\Delta E = 3\hbar\omega$ for the weakly coupled parameter set and $\Delta E = 20\hbar\omega$ for the high-temperature parameter set. The analytical solution is shown with a dashed line. The weakly coupled parameter set converges with four qubits corresponding to 16 energy levels. The high-temperature parameter set requires five qubits corresponding to 32 energy levels to represent the harmonic oscillator position sufficiently.

---

[*)]Contributed equally
[†)]Email: gsolomon@chem.ku.dk

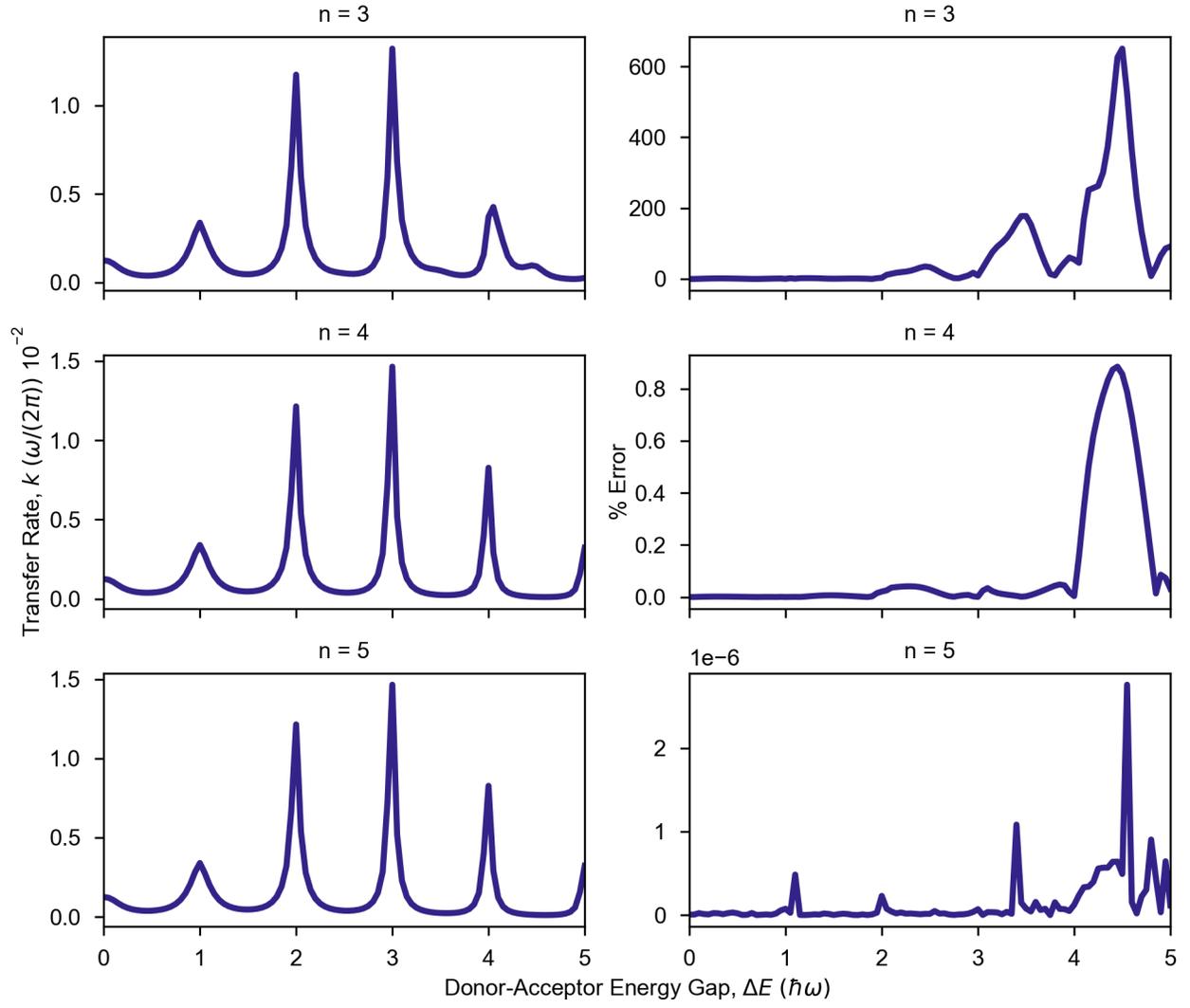

FIG. S2. **Harmonic Oscillator Truncation for the Weakly Coupled Parameter Set.** For $n = 3$ to 5 qubits, the electron transfer rates and the percentage error relative to $n = 6$ qubits are shown as a function of the donor-acceptor energy gap.

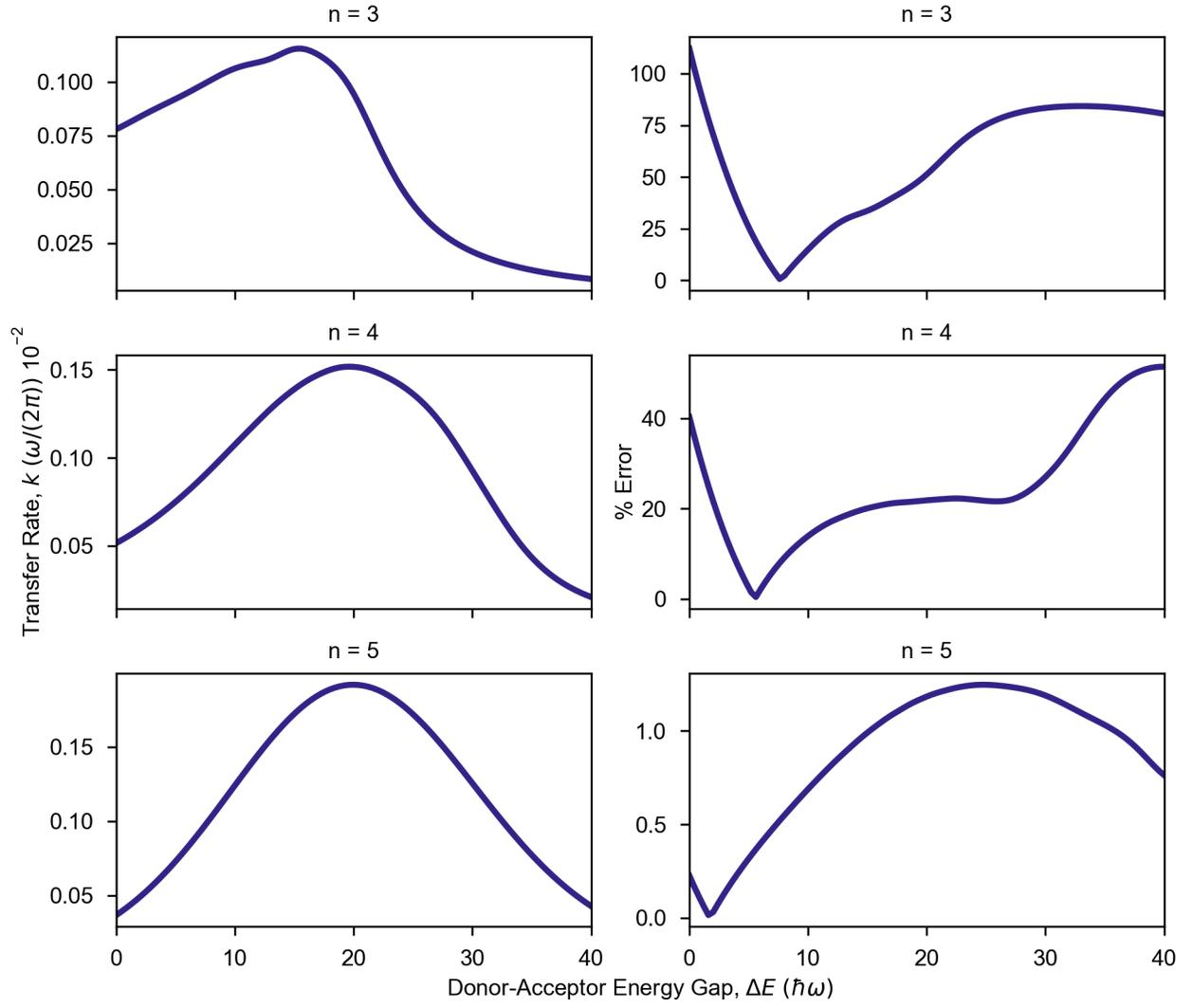

FIG. S3. **Harmonic Oscillator Truncation for the High-Temperature Parameter Set.** For $n = 3$ to $5$ qubits, the electron transfer rates and the percentage error relative to $n = 6$ qubits are shown as a function of the donor-acceptor energy gap.

## II. PRESERVATION OF NON-MARKOVIAN DYNAMICS IN THE REPEATED INTERACTION MODEL

We confirmed that the non-Markovian dynamics observed for the weakly coupled parameter set was preserved using the repeated interaction model by performing a simulation where the concurrence used in the Rivas, Huelga, Plenio measure was calculated and compared to the Lindblad result, see Figure S4.[1] We see the same increases in entanglement for the repeated interaction model, confirming that as the repeated interaction model converges to the Lindblad result, the backflow of information from the reaction coordinate to the electronic system is preserved.

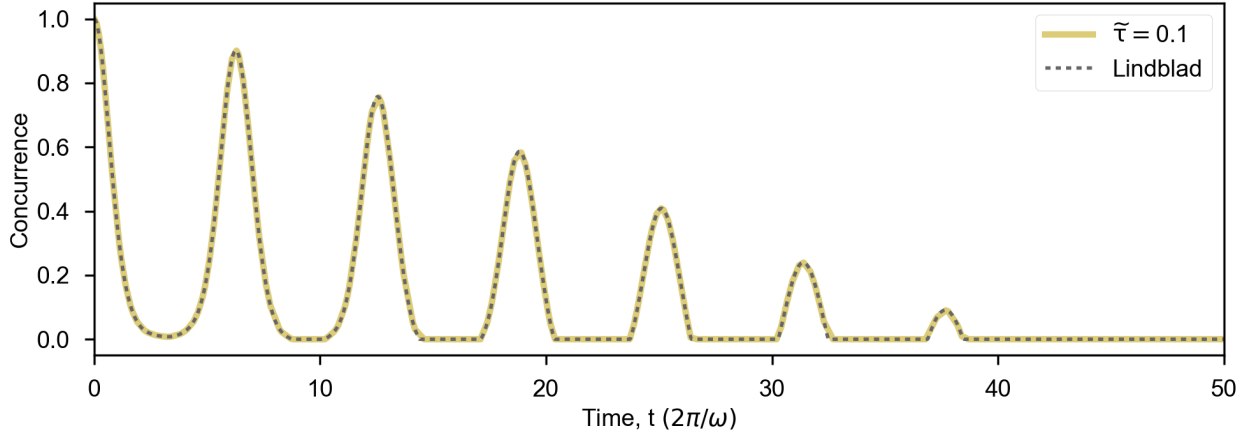

FIG. S4. **Concurrence Used in the Rivas, Huelga, Plenio Measure from a Repeated Interaction Simulation.** For the weakly coupled parameter set with $\Delta E = 3\hbar\omega$, the entanglement between the electron and a non-interacting ancilla (started in a maximally entangled state) is shown as a function of time. The concurrence does not monotonically decrease, showing non-Markovian dynamics. The repeated interaction simulation with $\tau = 0.1(2\pi/\omega)$ matches the Lindblad simulation result.

## III. COMPARISON TO ANALYTICAL BOUNDS

We would like to make a connection between our results and the analytical work of Pocrnic et al.[2] For example, it would be interesting to know if our result that the high-temperature parameter set converged with only 10 times as many repeated interactions as the weakly coupled parameter set would be expected or unexpected based on previously established theory.

In their work, Pocrnic et al. gave an analytical bound for the number of repeated interactions needed to converge to Lindblad dynamics (see Corollary 2.1 in Ref. 2), which for our system and setup reduces to

$$s \in O\left(\frac{t^2}{\epsilon}\left(\|\mathcal{L}\|_{1\to 1}^2 + \max\{\|H_{\mathrm{ET}}\|, \|H_{\mathrm{int}}\|\}^4\right)\right), \tag{1}$$

where $s$ is the number of repeated interactions needed to converge to Lindblad dynamics, $t$ is the total time of the simulation, and $\epsilon$ is the error. $\|\mathcal{L}\|_{1\to 1}$ is the induced Schatten $1 \to 1$ norm of the Lindbladian superoperator, defined as $\|\mathcal{L}\|_{1\to 1} = \max_A \|\mathcal{L}(A)\|_1$, where $\|A\|_1 = 1$. To see how our result compares to this asymptotic scaling result, we calculated the numerical value of $\|\mathcal{L}\|_{1\to 1}^2 + \max\{\|H_{\mathrm{ET}}\|, \|H_{\mathrm{int}}\|\}^4$ for each parameter set. It is difficult to calculate the induced superoperator norm exactly because it requires optimizing over all possible matrices with unit norm. To estimate this norm we perform Monte Carlo sampling over Hermitian matrices, with $\Delta E$ set to the highest value used for each parameter set. Our estimates are strictly less than the actual induced norm because there could be non-Hermitian matrices that give a larger value for $\|\mathcal{L}(A)\|_1$, meaning that we might underestimate the upper bound for the number of repeated interactions.

In Table S5 we show the ratios of the estimated values of $\|\mathcal{L}\|_{1\to 1}^2 + \max\{\|H_{\mathrm{ET}}\|, \|H_{\mathrm{int}}\|\}^4$ relative to the weakly coupled parameter set for the three other donor-acceptor parameter sets and compared them to the relative scaling we observed in our simulations. This comparison is only valid if the prefactors not captured by the Big O notation in Equation S5 are the same for all the parameter sets, and if we are in the limit of $\tau$ approaching zero. While we

suspect that these conditions may be satisfied, proving that this comparison is valid or making a more meaningful comparison would take further analytical work that is beyond the scope of this current study.

TABLE S5. **Required Number of Repeated Interactions Relative to Weakly Coupled Parameter Set.** The analytical values are based on asymptotic scaling laws without constant prefactors and should be interpreted qualitatively.

| Parameter set | Analytical | Observed |
|---|---|---|
| Strongly damped | 1.0032 | 1 |
| Strongly coupled | 1.0192 | 1 |
| High temperature | 1361.6 | 10 |



\end{document}